\documentclass[letterpaper]{jpconf}
\usepackage{amsfonts,amssymb}
\usepackage{graphicx,wrapfig}
\usepackage{booktabs}

\newcommand{\genasis}{{\sc GenASiS}}
\newcommand{\rShock}{R_{\mbox{\tiny Sh}}}
\newcommand{\rPNS}{R_{\mbox{\tiny PNS}}}
\newcommand{\model}[1]{\mbox{B}{#1}}
\newcommand{\kTurb}{k_{\mbox{\tiny T}}}
\newcommand{\lambdaTurb}{\lambda_{\mbox{\tiny T}}}
\newcommand{\EkinTurb}{E_{\mbox{\tiny k}}^{\mbox{\tiny T}}}
\newcommand{\lambdaMag}{\bar{\lambda}_{\mbox{\tiny m}}}
\newcommand{\timeAverage}[3]{\langle{#1}\rangle_{#2\hspace{0.05cm}\mbox{\tiny s}}^{#3\hspace{0.05cm}\mbox{\tiny s}}}
\newcommand{\vect}[1]{{\bf{#1}}}
\newcommand{\pderiv}[2]{\frac{{\partial #1}}{{\partial #2}}}
\newcommand{\f}[2]{\frac{#1}{#2}}
\newcommand{\volAvg}[3]{\langle{#1}\rangle_{#2}^{#3}}
\newcommand{\surfAvg}[3]{\langle{#1}\rangle_{#2}^{#3}}
\newcommand{\lineAvg}[3]{\langle{#1}\rangle_{#2}^{#3}}
\newcommand{\divergence}[1]{\nabla\cdot{#1}}
\newcommand{\gradient}[1]{\nabla{#1}}
\newcommand{\curl}[1]{\nabla\times{#1}}

\begin{document}

\title{Turbulent magnetic field amplification from spiral SASI modes in core-collapse supernovae}

\author{E Endeve$^{1}$, C Y Cardall$^{2,3}$, R D Budiardja$^{2,3,4}$ and A Mezzacappa$^{1,2,3}$}

\address{$^1$Computer Science and Mathematics Division, Oak Ridge National Laboratory (ORNL), Oak Ridge, TN 37831-6354, USA}
\address{$^2$Physics Division, ORNL, Oak Ridge, TN 37831, USA}
\address{$^3$Department of Physics and Astronomy, University of Tennessee, Knoxville, TN 37996, USA}
\address{$^4$Joint Institute for Heavy Ion Research, ORNL, Oak Ridge, TN 37831, USA}

\ead{endevee@ornl.gov}

\begin{abstract}
We describe the initial implementation of magnetohydrodynamics (MHD) in our astrophysical simulation code \genasis.  
Then, we present MHD simulations exploring the capacity of the stationary accretion shock instability (SASI) to generate magnetic fields by adding a weak magnetic field to an initially spherically symmetric fluid configuration that models a stalled shock in the post-bounce supernova environment.  
Upon perturbation and nonlinear SASI development, shear flows associated with the spiral SASI mode contributes to a widespread and turbulent field amplification mechanism.  
While the SASI may contribute to neutron star magnetization, these simulations do not show qualitatively new features in the global evolution of the shock as a result of SASI-induced magnetic field amplification.  
\end{abstract}

\section{Introduction}

Shortly after the discovery of pulsars \cite{hewish_etal_1968}, the potential role of magnetic fields in the core-collapse supernova (CCSN) explosion mechanism began to be investigated (e.g., \cite{leblancWilson_1970,bisnovatyi-kogan_etal_1976,meier_etal_1976,symbalisty_1984}).  
In principle, a differentially rotating proto-neutron star (PNS) could give rise to magnetically powered explosions.  
An early conclusion, however, was that (unrealistically) rapid rotation \emph{and} strong magnetic fields would be needed at the pre-collapse stage for magnetic fields to play a principal role in the explosion dynamics \cite{leblancWilson_1970,symbalisty_1984}.  

More recently, interest in strong magnetic fields has returned in connection with a number of observables related to CCSNe, including asymmetries in the explosion ejecta \cite{wheeler_etal_2002}, natal neutron star kick velocities \cite{laiQian_1998}, and the high-energy electromagnetic activity connected to some neutron stars known as magnetars, or Anomalous X-ray Pulsars (AXPs) and Soft Gamma Repeaters (SGRs) (e.g., \cite{duncanThompson_1992,thompsonDuncan_2001,hurley_etal_2005,woodsThompson_2006}).  
The theoretical discovery of the magneto-rotational instability (MRI) \cite{balbusHawley_1991} and its application to CCSNe \cite{akiyama_etal_2003} has also contributed to renewed interest in the possible role of magnetic fields in the explosion of some supernovae (i.e., those from rapidly rotating progenitor cores; e.g., \cite{wheeler_etal_2002,obergaulinger_etal_2005,moiseenko_etal_2006,burrows_etal_2007,takiwaki_etal_2009}). 
Still, magneto-rotationally driven CCSNe likely would be peculiar events, since magnetic progenitor cores tend to rotate slowly \cite{heger_etal_2005}.  

Leaving aside the explosion mechanism, the relationship between the formation of neutron star magnetic fields and CCSNe is still an open and interesting question, particularly in the case of AXPs and SGRs \cite{lorimerKramer_2005} (although magneto-rotational processes are likely involved \cite{thompsonDuncan_1993,bonanno_etal_2003}).  

The lack of sufficient rotational energy in magnetized pre-collapse progenitor cores, as predicted by stellar evolution models, has sparked some recent interest in MHD processes in non-rotating CCSN environments \cite{endeve_etal_2010,guilet_etal_2011,obergaulingerJanka_2011,endeve_etal_2012}.  
In particular, Endeve et al. \cite{endeve_etal_2010,endeve_etal_2012} studied magnetic field amplification by the stationary accretion shock instability (SASI \cite{blondin_etal_2003}).  
The SASI may play an important role in neutrino-powered explosions \cite{bruenn_etal_2006,buras_etal_2006,mezzacappa_etal_2007,marekJanka_2009}.  
Thus, magnetic fields may be an important part of a supernova model if the SASI is found to be sensitive to their presence.  

This paper complements the investigations initiated in \cite{endeve_etal_2010}.  
We begin with a description of the numerical methods used (the initial implementation of MHD in our astrophysical simulation code \genasis), including results from some standard tests.  
Then, with a new set of high-resolution simulations, we investigate the growth and impact of magnetic fields during operation of the SASI.  
We also attempt to quantify the levels of neutron star magnetization that may result from SASI dynamics.  
We find that the magnetic energy may grow exponentially in turbulent flows driven by the spiral SASI mode.  
(The magnetic energy growth time, based on the turnover time of SASI-driven turbulence, is estimated to be a few milliseconds.)  
The presence of amplified magnetic fields results in less kinetic energy on small spatial scales, but we find no impact of magnetic fields on global shock dynamics.  
However, magnetic field evolution remains sensitive to numerical resolution.  
We argue that MHD processes associated with the SASI may contribute significantly to strong, small-scale PNS magnetic fields, and provide a connection between the magnetic fields of neutron stars at birth and supernova dynamics.  
The saturation energies may be sufficient to power flaring activity of AXPs, and possibly SGRs.  
Moreover, their formation does not require progenitor rotation.  

\section{Equations and numerical methods}

To study magnetic field amplification from SASI-driven flows we solve the equations of ideal MHD.  
The fluid mass, momentum and energy densities obey conservation laws with sources
\begin{equation}
  \pderiv{\vect{U}}{t}+\divergence{\vect{F}}=\vect{S}, \label{eq:conservationLaws}
\end{equation}
and the evolution of the magnetic field is governed by Faraday's law (the induction equation)
\begin{equation}
  \pderiv{\vect{B}}{t}=-\curl{\vect{E}}, \label{eq:faradayLaw}
\end{equation}
where the vector of conserved variables is $\vect{U}=[\rho,\rho\vect{u},E]^{T}$, the vector of fluxes is $\vect{F}=[\rho\vect{u},\rho\vect{uu}+\vect{I}P^{\star}-\vect{BB},(E+P^{\star})\vect{u}-\vect{B}(\vect{B}\cdot\vect{u})]^{T}$, and the vector of sources is $\vect{S}=[0,-\rho\gradient{\Phi},\rho\vect{u}\cdot\gradient{\Phi}]^{T}$.  
The electric field is $\vect{E}=-\vect{u}\times\vect{B}$ (assuming a perfectly conducting fluid).  
We denote the combined vector of evolved variables by $\vect{W}=[\vect{U},\vect{B}]^{T}$.  
(Thus, $\vect{F}=\vect{F}(\vect{W})$ and $\vect{E}=\vect{E}(\vect{W})$.)
Variables $\rho$, $\vect{u}$, $E=e+\rho u^{2}/2+B^{2}/2$, $\vect{B}$ denote mass density, fluid velocity, fluid energy density, and magnetic field, respectively.  
(We adopt units where the vacuum permeability is $\mu_{0}=1$.)  
The total pressure (thermal plus magnetic) is $P^{\star}=P+B^{2}/2$.  
In this study, the internal energy density is related to the pressure via the ideal gas law $e=P/(\gamma-1)$, where $\gamma$ is the adiabatic exponent (ratio of specific heats).  

We report on simulations of flows around a compact object (the PNS), and we take the gravitational potential to be given by the point-mass formula $\Phi=-GM/r$, where $G$ is Newton's constant, $M$ the mass of the central object, and $r$ the radial distance from the center.  

\subsection{Magnetohydrodynamics scheme in \genasis}
\label{sec:mhdScheme}

We adopt an explicit finite volume method (e.g., \cite{leveque_2002}) for the simultaneous time-integration of Eqs. (\ref{eq:conservationLaws}) and (\ref{eq:faradayLaw}).  
Cartesian coordinates are used to represent points in a three-dimensional computational domain with sides $L_{x}$, $L_{y}$, and $L_{z}$, and volume $V$.  
The computational domain is subdivided into $N_{x}\times N_{y}\times N_{z}$ computational cells with sides $\Delta x$, $\Delta y$, and $\Delta z$, and volume $\Delta V$.  
The $x$-coordinate of the interface separating cells indexed $i$ and $i+1$ is denoted $x_{i+1/2}=x_{1/2}+i\,\Delta x$, with $i=0,\ldots,N_{x}$, and the corresponding cell center coordinate is $x_{i}=(x_{i-1/2}+x_{i+1/2})$/2.  
Similar definitions apply for the other coordinate dimensions.  
The time step $\Delta t$ increments time from $t^{n}$ to $t^{n+1}=t^{n}+\Delta t$.  

Integrating Eq. (\ref{eq:conservationLaws}) over the spacetime volume element $\Delta V\times\Delta t$, applying Gauss' theorem, and representing time integrals of the fluxes and sources with the rectangle rule results in the time-explicit finite volume update formula
\begin{eqnarray}
  \volAvg{\vect{U}}{ijk}{n+1}
  &=&
  \volAvg{\vect{U}}{ijk}{n}
  -\f{\Delta t}{\Delta x}\left[\surfAvg{\vect{F}_{x}}{i+\f{1}{2}jk}{n}-\surfAvg{\vect{F}_{x}}{i-\f{1}{2}jk}{n}\right]
  -\f{\Delta t}{\Delta y}\left[\surfAvg{\vect{F}_{y}}{ij+\f{1}{2}k}{n}-\surfAvg{\vect{F}_{y}}{ij-\f{1}{2}k}{n}\right] 
  \nonumber \\
  & &
  -\f{\Delta t}{\Delta z}\left[\surfAvg{\vect{F}_{z}}{ijk+\f{1}{2}}{n}-\surfAvg{\vect{F}_{z}}{ijk-\f{1}{2}}{n}\right]
  +\Delta t\,\volAvg{\vect{S}}{ijk}{n}. \label{eq:finiteVolume}
\end{eqnarray}
Angle brackets on $\vect{U}$ and $\vect{S}$ (centered on the geometric centers of computational cells) denote volume averages, while angle brackets on the fluxes (centered on the faces of computational cells) denote area averages.  
We use the HLL Riemann solver \cite{harten_etal_1983} to compute the interface fluxes, given left and right states $\vect{U}^{\mbox{\tiny L}}$ and $\vect{U}^{\mbox{\tiny R}}$, respectively.  
The structure between the fastest left and right propagating waves (fast magnetosonic waves with propagation speeds denoted $-\alpha^{-}$ and $\alpha^{+}$, respectively) in the so-called Riemann fan is then represented by a single average state $\vect{U}^{\star}$.  
The HLL flux (for example in the $x$-direction $\vect{F}_{x}^{\star}$) is obtained from the Rankine-Hugoniot jump conditions cross the waves separating the left and right states: $-\alpha_{x}^{-}(\vect{U}^{\mbox{\tiny L}}-\vect{U}^{\star})=(\vect{F}_{x}^{\mbox{\tiny L}}-\vect{F}_{x}^{\star})$ and $\alpha_{x}^{+}(\vect{U}^{\star}-\vect{U}^{\mbox{\tiny R}})=(\vect{F}_{x}^{\star}-\vect{F}_{x}^{\mbox{\tiny R}})$.  
The resulting flux is used in Eq. (\ref{eq:finiteVolume}); i.e., 
\begin{equation}
  \surfAvg{\vect{F}_{x}}{}{n}=
  \f{\alpha_{x}^{+}\vect{F}_{x}^{n,\mbox{\tiny L}}+\alpha_{x}^{-}\vect{F}_{x}^{n,\mbox{\tiny R}}}
  {\alpha_{x}^{+}+\alpha_{x}^{-}}
  -\f{\alpha_{x}^{+}\alpha_{x}^{-}}{\alpha_{x}^{+}+\alpha_{x}^{-}}
  \left[\vect{U}^{n,\mbox{\tiny R}}-\vect{U}^{n,\mbox{\tiny L}}\right], \label{eq:HLLFlux}
\end{equation}
where the fluxes $\vect{F}_{x}^{n,\mbox{\tiny L(R)}}$ are evaluated from $\vect{W}^{n,\mbox{\tiny L(R)}}$.  
Similar expressions are obtained for the other flux components.  
The wave speeds are computed from the left and right states of the interface \cite{davis_1988} $\alpha_{x}^{\pm}=\max[0,\pm\lambda_{x}^{\pm}(\vect{W}^{\mbox{\tiny L}}),\pm\lambda_{x}^{\pm}(\vect{W}^{\mbox{\tiny R}})]$, where $\lambda_{x}^{\pm}=u_{x}\pm c_{f,x}$ and $c_{f,x}$ is the fast magnetosonic speed.  
The HLL Riemann solver is among the simplest approximate Riemann solvers available for MHD.  
It relies on minimal information about the characteristic structure of the underlying hyperbolic system (only two eigenvalues of the flux Jacobian matrix).  
As such, the HLL flux has been frequently used to construct simple, efficient, and robust solvers for hyperbolic systems, including classical MHD \cite{londrilloDelZanna_2004,ziegler_2004}, and special \emph{and} general relativistic MHD \cite{delZanna_etal_2003,gammie_etal_2003,duez_etal_2005,delZanna_etal_2007}.  
The averaging over the Riemann fan results in diffusive evolution of intermediate waves (e.g., contact and Alfv{\'e}n).  
However, the intermediate waves can be restored in the HLL framework \cite{toro_etal_1994,linde_2002,gurski_2004,li_2005,miyoshiKusano_2005,mignone_etal_2009}.  

For second-order spatial accuracy the left and right states needed by the Riemann solver are reconstructed from cell-centered volume averages with monotonic linear interpolation: $\vect{U}_{i+1/2}^{n,\mbox{\tiny L}}=\volAvg{\vect{U}}{i}{n}+\mathcal{D}_{x}[\volAvg{\vect{U}}{i}{n}](x_{i+1/2}-x_{i})$ and $\vect{U}_{i+1/2}^{n,\mbox{\tiny R}}=\vect{U}_{i+1}^{n}-\mathcal{D}_{x}[\vect{U}_{i+1}^{n}](x_{i+1}-x_{i+1/2})$.  
Monotonic interpolation of a variable $f_{i}^{n}$ is achieved with the the generalized minmod slope (e.g., \cite{kurganovTadmor_2000})
\begin{equation}
  \mathcal{D}_{x}\left[f_{i}^{n}\right]
  =\Delta x^{-1}\,\mbox{minmod}
  \left[
    \vartheta\left(f_{i}^{n}-f_{i-1}^{n}\right),\, 
    0.5\left(f_{i+1}^{n}-f_{i-1}^{n}\right),\, 
    \vartheta\left(f_{i+1}^{n}-f_{i}^{n}\right)
  \right], 
\end{equation}
where $\vartheta\in[1,2]$.  
We use $\vartheta=1.4$ in the simulations presented in this paper.  

The magnetic field is evolved with the constrained transport (CT) method of Evans \& Hawley \cite{evansHawley_1988}.  
Magnetic field components are collocated on the faces of computational cells.  
Integrating the induction equation (Eq. \ref{eq:faradayLaw}) over cell faces and $\Delta t$, applying Stoke's theorem, and replacing time integrals of electric field components with the rectangle rule results in time-explicit update formulae for area averaged magnetic field components
\begin{eqnarray}
  \surfAvg{B_{x}}{i+\f{1}{2}jk}{n+1}
  &=&\surfAvg{B_{x}}{i+\f{1}{2}jk}{n} \nonumber \\
  & &
  +\f{\Delta t}{\Delta z}\left[\lineAvg{E_{y}}{i+\f{1}{2}jk+\f{1}{2}}{n}-\lineAvg{E_{y}}{i+\f{1}{2}jk-\f{1}{2}}{n}\right]
  -\f{\Delta t}{\Delta y}\left[\lineAvg{E_{z}}{i+\f{1}{2}j+\f{1}{2}k}{n}-\lineAvg{E_{z}}{i+\f{1}{2}j-\f{1}{2}k}{n}\right]
  \label{eq:inductionBx} \\
  \surfAvg{B_{y}}{ij+\f{1}{2}k}{n+1}
  &=&\surfAvg{B_{y}}{ij+\f{1}{2}k}{n} \nonumber \\
  & &
  +\f{\Delta t}{\Delta x}\left[\lineAvg{E_{z}}{i+\f{1}{2}j+\f{1}{2}k}{n}-\lineAvg{E_{z}}{i-\f{1}{2}j+\f{1}{2}k}{n}\right]
  -\f{\Delta t}{\Delta z}\left[\lineAvg{E_{x}}{ij+\f{1}{2}k+\f{1}{2}}{n}-\lineAvg{E_{x}}{ij+\f{1}{2}k-\f{1}{2}}{n}\right]
  \label{eq:inductionBy} \\
  \surfAvg{B_{z}}{ijk+\f{1}{2}}{n+1}
  &=&\surfAvg{B_{z}}{ijk+\f{1}{2}}{n} \nonumber \\
  & &
  +\f{\Delta t}{\Delta y}\left[\lineAvg{E_{x}}{ij+\f{1}{2}k+\f{1}{2}}{n}-\lineAvg{E_{x}}{ij-\f{1}{2}k+\f{1}{2}}{n}\right]
  -\f{\Delta t}{\Delta x}\left[\lineAvg{E_{y}}{i+\f{1}{2}jk+\f{1}{2}}{n}-\lineAvg{E_{y}}{i-\f{1}{2}jk+\f{1}{2}}{n}\right].
  \label{eq:inductionBz}
\end{eqnarray}
Angle brackets on electric field components (centered on the edges of computational cells) denote line averages.  
Note that with the divergence of the magnetic field defined by
\begin{equation}
  \volAvg{\divergence{\vect{B}}}{ijk}{}
  =
  \f{\surfAvg{B_{x}}{i+\f{1}{2}jk}{}-\surfAvg{B_{x}}{i-\f{1}{2}jk}{}}{\Delta x}
  +\f{\surfAvg{B_{y}}{ij+\f{1}{2}k}{}-\surfAvg{B_{y}}{ij-\f{1}{2}k}{}}{\Delta y}
  +\f{\surfAvg{B_{z}}{ijk+\f{1}{2}}{}-\surfAvg{B_{z}}{ijk-\f{1}{2}}{}}{\Delta z}, 
\end{equation}
the magnetic field update given by Eqs. (\ref{eq:inductionBx})-(\ref{eq:inductionBz}) results in $\volAvg{\divergence{\vect{B}}}{ijk}{n+1}=\volAvg{\divergence{\vect{B}}}{ijk}{n}$---independent of how the electric fields are computed.  
Thus, the magnetic field remains divergence-free, provided it is so initially.  

We use HLL-type expressions \cite{londrilloDelZanna_2004} (see also \cite{kurganov_etal_2001}) to compute edge centered electric fields in Eqs. (\ref{eq:inductionBx})-(\ref{eq:inductionBz}).  
The $z$-component centered on, say, ($x_{i+1/2},y_{j+1/2},z_{k}$) is
\begin{eqnarray}
  \lineAvg{E_{z}}{}{n}
  &=&
  \f{\alpha_{x}^{+}\alpha_{y}^{+}E_{z}^{n,\mbox{\tiny SW}}
     +\alpha_{x}^{+}\alpha_{y}^{-}E_{z}^{n,\mbox{\tiny NW}}
     +\alpha_{x}^{-}\alpha_{y}^{+}E_{z}^{n,\mbox{\tiny SE}}
     +\alpha_{x}^{-}\alpha_{y}^{-}E_{z}^{n,\mbox{\tiny NE}}}
     {(\alpha_{x}^{+}+\alpha_{x}^{-})(\alpha_{y}^{+}+\alpha_{y}^{-})} \nonumber \\
     & &
     +\f{\alpha_{x}^{+}\alpha_{x}^{-}}{(\alpha_{x}^{+}+\alpha_{x}^{-})}\left[B_{y}^{n,\mbox{\tiny E}}-B_{y}^{n,\mbox{\tiny W}}\right]
     -\f{\alpha_{y}^{+}\alpha_{y}^{-}}{(\alpha_{y}^{+}+\alpha_{y}^{-})}\left[B_{x}^{n,\mbox{\tiny N}}-B_{x}^{n,\mbox{\tiny S}}\right], 
     \label{eq:HLLField}
\end{eqnarray}
where $\alpha_{x}^{\pm}$ and $\alpha_{y}^{\pm}$ are computed by taking the maximum among the four states surrounding the edge.  
The HLL electric field in Eq. (\ref{eq:HLLField}) considers the four cells sharing the edge where the electric field is to be evaluated.  
It is derived from the principles used in the derivation of the HLL flux in Eq. (\ref{eq:HLLFlux}), but applied in two dimensions.  
For the electric field centered on ($x_{i+1/2},y_{j+1/2},z_{k}$), superscripts $\mbox{SW}$, $\mbox{NW}$, $\mbox{SE}$, and $\mbox{NE}$ indicate that a quantity is obtained from values in cells with centers ($x_{i},y_{j},z_{k}$), ($x_{i},y_{j+1},z_{k}$), ($x_{i+1},y_{j},z_{k}$), and ($x_{i+1},y_{j+1},z_{k}$), respectively.  
Magnetic field components are centered on cell faces, and the cells with centers ($x_{i},y_{j},z_{k}$) and ($x_{i},y_{j+1},z_{k}$) share $\surfAvg{B_{y}}{ij+1/2k}{}$, which is labeled with superscript $\mbox{W}$.  
Similarly, the cells with centers ($x_{i},y_{j},z_{k}$) and ($x_{i+1},y_{j},z_{k}$) share $\surfAvg{B_{x}}{i+1/2jk}{}$, which is labeled with superscript $\mbox{S}$.  
Velocity components are cell centered and must be interpolated in two dimensions, while face centered magnetic field components are interpolated in one dimension only.  
The last two terms in Eq. (\ref{eq:HLLField}) result from averaging over Riemann fans and introduce explicit dissipation of the magnetic field.  
They act to stabilize the time evolution in under-resolved regions, and can become dominant if the bulk of the magnetic fields evolve on spatial scales comparable to the grid spacing (see Section \ref{sec:results}).  

The update formulae given by Eq. (\ref{eq:finiteVolume}) and Eqs. (\ref{eq:inductionBx})-(\ref{eq:inductionBz}) are equivalent to the forward Euler method and result in first-order temporal accuracy.  
Second-order accuracy in time is achieved with a total variation diminishing (TVD) Runge-Kutta method (e.g., \cite{shu_1997})
\begin{eqnarray}
  \volAvg{\vect{W}}{}{\star}
  &=&\volAvg{\vect{W}}{}{n}
  +\mathcal{L}\left[\volAvg{\vect{W}}{}{n}\right] \nonumber \\
  \volAvg{\vect{W}}{}{n+1}
  &=&
  \volAvg{\vect{W}}{}{n}
  +\left(\mathcal{L}\left[\volAvg{\vect{W}}{}{n}\right]+\mathcal{L}\left[\volAvg{\vect{W}}{}{\star}\right]\right)/2, \label{eq:RK2}
\end{eqnarray}
which is a convex combination of two forward Euler steps.  
The spatial operators and sources are here represented by $\mathcal{L}[\volAvg{\vect{W}}{}{}]$.  
The time step is determined from the Courant-Friedrichs-Lewy (CFL) condition: $\Delta t=\mbox{C}\times\min\left[\Delta t_{x},\Delta t_{y},\Delta t_{z}\right]$, where, for example, $\Delta t_{x}=\Delta x/\max[\alpha_{x}^{+},\alpha_{x}^{-}]$, $\mbox{C}\le1$ is the Courant number, and the minimum is taken among all computational cells in $V$.  

We point out that our MHD scheme is similar to the MC-HLL-UCT scheme described in \cite{londrilloDelZanna_2004} and the MHD scheme in the NIRVANA code \cite{ziegler_2004}.  (See also the semidiscrete central-upwind schemes developed for hyperbolic conservation laws and Hamilton-Jacobi equations in \cite{kurganov_etal_2001}.)  

\subsubsection{Numerical tests}

To demonstrate the correctness of our implementation of the MHD solver we present results from two well-known test problems, the circularly polarized Alfv{\'e}n wave and the Orszag-Tang vortex, computed in two spatial dimensions.  
Both tests use periodic boundary conditions everywhere.  
The Courant number is set to $\mbox{C}=0.4$, the slope limiting parameter is set to $\vartheta=1.4$, and adiabatic exponent $\gamma=5/3$.  

\underline{\emph{Circularly polarized (CP) Alfv{\'e}n wave}}
This test problem has been used by many authors (e.g., \cite{toth_2000,londrilloDelZanna_2004,stone_etal_2008}).  
In particular, T{\'o}th \cite{toth_2000} used this test in an extensive comparison study of multidimensional MHD schemes.  
The CP Alfv{\'e}n wave is an analytic nonlinear solution to the MHD equations.  
As such it provides a means to verify the formal order of the scheme through convergence testing.  
The wave propagates with an angle $\alpha=\tan^{-1}(2)\approx63.4^{\circ}$ relative to the $x$-axis.  
The problem is set up with constant background density $\rho_{0}=1$ and pressure $P_{0}=0.1$.  
The velocity (magnetic field) components are $u_{x}(B_{x})=u_{\parallel}(B_{\parallel})\cos\alpha-u_{\perp}(B_{\perp})\sin\alpha$, $u_{y}(B_{y})=u_{\parallel}(B_{\parallel})\sin\alpha+u_{\perp}(B_{\perp})\cos\alpha$, and $u_{z}(B_{z})=A\sin(2\pi x_{\parallel})$, with $u_{\perp}=B_{\perp}=A\sin(2\pi x_{\parallel})$ and $x_{\parallel}=x\cos\alpha+y\sin\alpha$.  
The background magnetic field parallel to the direction of wave propagation is $B_{\parallel}=1$ (i.e., the Alfv{\'e}n speed $v_{\mbox{\tiny A}}$ is unity).  
The parallel velocity $u_{\parallel}$ is set to zero for the traveling wave (TW) test and $-1$ for the standing wave (SW) test.  
The wave amplitude is $A=0.01$.  
Periodic boundary conditions can be used if the domain is $[x,y]\in[0,1/\cos\alpha]\times[0,1/\sin\alpha]$ ($L_{x}/L_{y}=2$, and we use $N_{x}=2N_{y}$ for square cells).  

\begin{table}
  \begin{center}
  \caption{$L_{1}$-error and convergence rate for the circularly polarized Alfv{\'e}n wave test.  \label{tab:convergenceResultsCPAW}}
  \begin{tabular}{ccccccccc}
    \midrule
    & \multicolumn{2}{c}{TW ($t_{f}=1$)} & 
    \multicolumn{2}{c}{SW ($t_{f}=1$)} &
    \multicolumn{2}{c}{TW ($t_{f}=5$)} &
    \multicolumn{2}{c}{SW ($t_{f}=5$)} \\
    \cmidrule(r){2-3} \cmidrule(r){4-5} \cmidrule(r){6-7} \cmidrule(r){8-9}
    $N_{x}$ & $L_{1}(B_{z})$ & Rate & $L_{1}(B_{z})$ & Rate & $L_{1}(B_{z})$ & Rate & $L_{1}(B_{z})$ & Rate \\
    \midrule
    32     & 1.373E-01 &         $-$ & 2.279E-01 &         $-$ & 4.974E-01 &         $-$ & 7.361E-01 &         $-$ \\
    64     & 5.196E-02 & $-1.40$ & 8.629E-02 & $-1.40$ & 1.395E-01 & $-1.83$ & 2.144E-01 & $-1.78$ \\
    128   & 1.433E-02 & $-1.86$ & 2.535E-02 & $-1.77$ & 5.053E-02 & $-1.46$ & 8.700E-02 & $-1.30$ \\
    256   & 3.625E-03 & $-1.98$ & 6.842E-03 & $-1.89$ & 1.483E-02 & $-1.77$ & 2.754E-02 & $-1.66$ \\
    512   & 9.085E-04 & $-2.00$ & 1.760E-03 & $-1.96$ & 3.945E-03 & $-1.91$ & 7.782E-03 & $-1.82$ \\
    1024 & 2.242E-04 & $-2.02$ & 4.491E-04 & $-1.97$ & 1.015E-03 & $-1.96$ & 2.059E-03 & $-1.92$ \\
    \midrule
  \end{tabular}
  \end{center}
\end{table}
Convergence results from the CP Alfv{\'e}n wave test (TW and SW tests), run for one ($t=1$) and five ($t=5$) grid crossings, are listed in Table \ref{tab:convergenceResultsCPAW}.  
The $L_{1}$-error norm is $L_{1}(B_{z})=\sum|B_{z}(t=0)-B_{z}(t=t_{f})|/\sum |B_{z}(t=0)|$, where the sums extend over all cells.  
The tests demonstrate second-order convergence with increasing resolution.  
The SW errors are larger (almost a factor 2) than the TW errors.  
(The convergence rate is given by $\log(L_{1}(B_{z},N_{x})/L_{1}(B_{z},N_{x}/2))/\log(2)$.)  

\underline{\emph{Orszag-Tang Vortex}}
This is another test which has be used extensively to benchmark multidimensional MHD codes (e.g., \cite{londrilloDelZanna_2004,stone_etal_2008}).  
Following the initialization in \cite{stone_etal_2008}, the computational domain is the unit square $[x,y]\in[0,1]\times[0,1]$, where the density and pressure are initially uniform $\rho=\f{25}{36\pi}$ and $P=\f{5}{12\pi}$ (resulting in sound speed $c_{\mbox{\tiny S}}=1$).  
The nonzero components of the velocity field are $u_{x}=-\sin(2\pi y)$ and $u_{y}=\sin(2\pi x)$, and the magnetic field is given by $B_{x}=-B_{0}\sin(2\pi y)$ and $B_{y}=B_{0}\sin(4\pi x)$ with $B_{0}=1/\sqrt{4\pi}$.  

\begin{figure}
  \includegraphics[angle=00,height=6.0354in,width=6.3in]{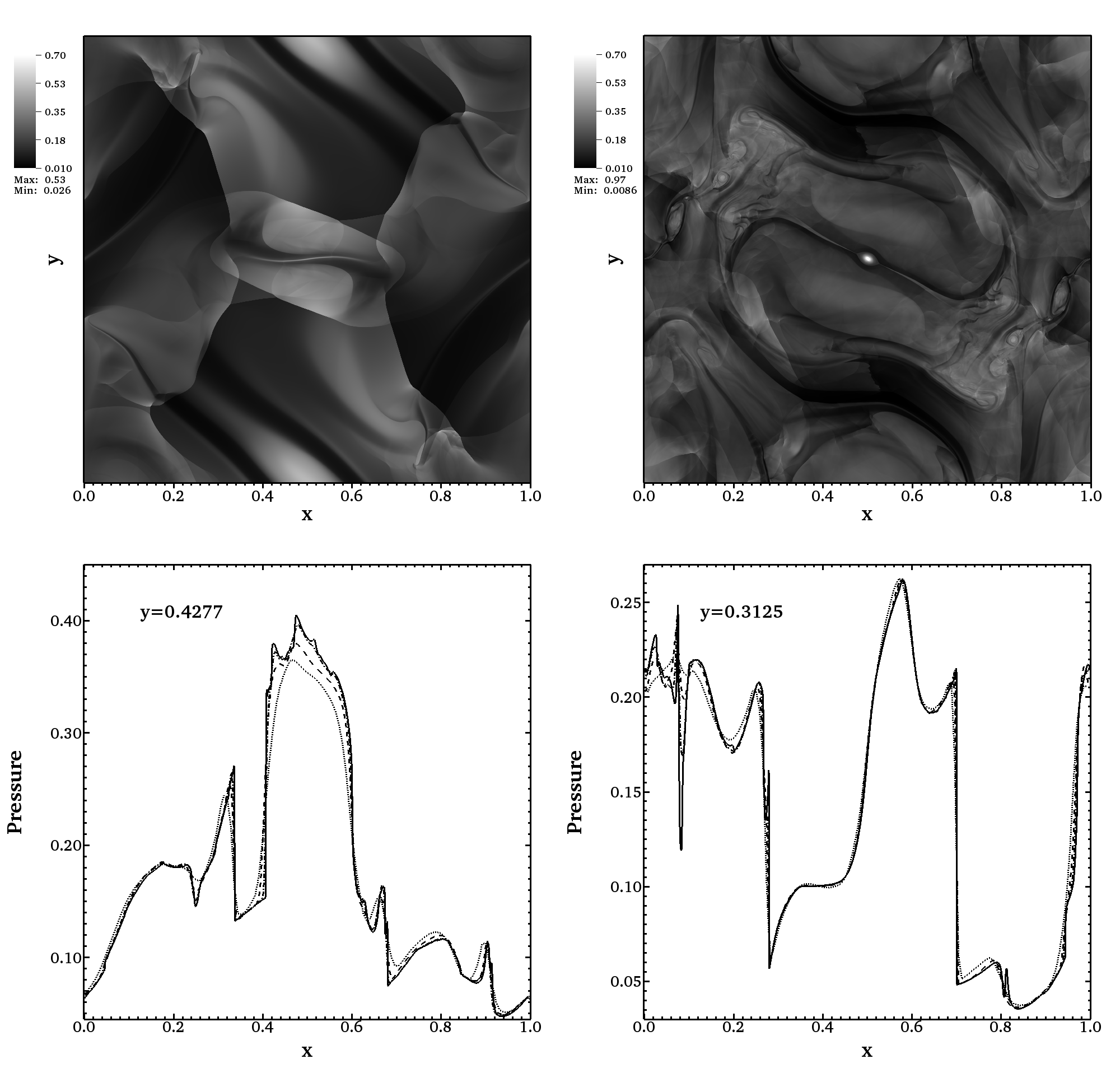}
    \caption{Results from running the Orszag-Tang vortex test with \genasis.  The upper panels shows the fluid pressure at times $t=0.5$ (left) and $t=1.0$ (right) for a model computed with $4096^{2}$ cells.  In the two lower panels we plot the thermal pressure versus $x$ at $t=0.5$ for $y=0.4277$ (left) and $y=0.3125$ (right) (cf. \cite{stone_etal_2008}).  Profiles are plotted for runs with varying spatial resolutions: $N_{x}\times N_{y}=128^{2}$ ($\dotted$), $256^{2}$ ($\dashed$), $512^{2}$ ($\chain$), and $4096^{2}$ ($\full$).  \label{fig:orszagTang}}
\end{figure}

Shocks develop quickly in the flow for $t>0$.  
The flow becomes very complex and develops MHD turbulence after multiple shock-shock interactions.  
Figure \ref{fig:orszagTang} displays results from the Orszag-Tang test computed with \genasis.  
With the profiles displayed in the lower panels, we have computed the $L_{1}$-error norm of the lower-resolution runs with respect to the high-resolution run ($4096^{2}$).  
The error norm decreases with increasing resolution at the first order rate (as expected for flows containing discontinuities).  

\subsection{Gravitational source terms}
\label{sec:gravitySources}

The gravitational force and power must be specified for problems involving gravity.  
In particular, constructing source terms which result in good energy conservation properties is of interest.  
Multiplication of the discrete mass conservation equation in Eq. (\ref{eq:finiteVolume}) with the potential $\volAvg{\Phi}{ijk}{}$ results in (after some algebra and assuming a static gravitational potential)
\begin{eqnarray}
  \volAvg{E_{g}}{ijk}{n+1}
  &=&\volAvg{E_{g}}{ijk}{n}
  -\f{\Delta t}{\Delta x}\left[\surfAvg{F_{x}^{E_{g}}}{i+\f{1}{2}jk}{n}-\surfAvg{F_{x}^{E_{g}}}{i-\f{1}{2}jk}{n}\right]
  -\f{\Delta t}{\Delta y}\left[\surfAvg{F_{y}^{E_{g}}}{ij+\f{1}{2}k}{n}-\surfAvg{F_{y}^{E_{g}}}{ij-\f{1}{2}jk}{n}\right]
  \nonumber \\
  & &
  -\f{\Delta t}{\Delta z}\left[\surfAvg{F_{z}^{E_{g}}}{ijk+\f{1}{2}}{n}-\surfAvg{F_{z}^{E_{g}}}{ijk-\f{1}{2}}{n}\right]
  +\f{\Delta t}{2}\left[\surfAvg{F_{x}^{\rho}\,g_{x}}{i-\f{1}{2}jk}{n}+\surfAvg{F_{x}^{\rho}\,g_{x}}{i+\f{1}{2}jk}{n}\right]
  \nonumber \\
  & &
  +\f{\Delta t}{2}\left[\surfAvg{F_{y}^{\rho}\,g_{y}}{ij-\f{1}{2}k}{n}+\surfAvg{F_{y}^{\rho}\,g_{y}}{ij+\f{1}{2}k}{n}\right]
  +\f{\Delta t}{2}\left[\surfAvg{F_{z}^{\rho}\,g_{z}}{ijk-\f{1}{2}}{n}+\surfAvg{F_{z}^{\rho}\,g_{z}}{ijk+\f{1}{2}}{n}\right], 
  \label{eq:gravitationalEnergyFiniteVolume}
\end{eqnarray}
where the gravitational energy density is $\volAvg{E_{g}}{ijk}{n}=\volAvg{\rho}{ijk}{n}\volAvg{\Phi}{ijk}{}$ and the $x$-component of the gravitational energy flux is $\surfAvg{F_{x}^{E_{g}}}{i+1/2jk}{n}=\surfAvg{F_{x}^{\rho}}{i+1/2jk}{n}\,(\volAvg{\Phi}{ijk}{}+\volAvg{\Phi}{i+1jk}{})/2$.  
We have also defined $\surfAvg{F_{x}^{\rho}\,g_{x}}{i+1/2jk}{n}=\surfAvg{F_{x}^{\rho}}{i+1/2jk}{n}(\volAvg{\Phi}{i+1jk}{}-\volAvg{\Phi}{ijk}{})/\Delta x$.  
(The $x$-component of the mass flux in Eq. (\ref{eq:finiteVolume}) is denoted $\surfAvg{F_{x}^{\rho}}{i+1/2jk}{n}$.)  
The last three terms in Eq. (\ref{eq:gravitationalEnergyFiniteVolume}) suggest a finite volume representation of the gravitational power $\volAvg{S^{E}}{ijk}{n}=-\volAvg{\rho\vect{u}\cdot\gradient{\Phi}}{ijk}{n}$.  
This choice results in energy conservation to machine precision; i.e., 
\begin{equation}
  \mathcal{E}^{n+1}
  =\sum_{i=1}^{N_{x}}\sum_{j=1}^{N_{y}}\sum_{k=1}^{N_{z}}\left(\volAvg{E}{ijk}{n+1}+\volAvg{E_{g}}{ijk}{n+1}\right)\Delta V
  =\mathcal{E}^{n}+\mbox{boundary energy fluxes}.  
\end{equation}
This result relies on the assumption of a static gravitational potential, and is sufficient for the simulations presented in this paper.  
We will present a suitable generalized discretization of the gravitational power---valid for time-dependent potentials---in a forthcoming paper.  

A finite volume representation of the $x$-component of the gravitational force is given by
\begin{equation}
  \volAvg{S^{\rho u_{x}}}{ijk}{n}=
  -\f{1}{2}\left[\surfAvg{\rho g_{x}}{i-\f{1}{2}jk}{n}+\surfAvg{\rho g_{x}}{i+\f{1}{2}jk}{n}\right],
\end{equation}
where $\surfAvg{\rho g_{x}}{i+1/2jk}{n}=\surfAvg{\rho}{i+1/2jk}{n}(\volAvg{\Phi}{i+1jk}{}-\volAvg{\Phi}{ijk}{})/\Delta x$.  
(Similar expressions are used for the $y$- and $z$-components.)  
For the density centered on $x=x_{i+1/2}$ we use the so-called HLL average
\begin{equation}
  \surfAvg{\rho}{i+\f{1}{2}}{}
  = \f{\alpha_{x,i+\f{1}{2}}^{+}\rho_{i+\f{1}{2}}^{\mbox{\tiny R}}+\alpha_{x,i+\f{1}{2}}^{-}\rho_{i+\f{1}{2}}^{\mbox{\tiny L}}
        -[F_{x,i+\f{1}{2}}^{\rho,\mbox{\tiny R}}-F_{x,i+\f{1}{2}}^{\rho,\mbox{\tiny L}}]}
     {\alpha_{x,i+\f{1}{2}}^{+}+\alpha_{x,i+\f{1}{2}}^{-}} \label{eq:hllAvgDensity}.
\end{equation}
Note that the expression for the gravitational power does \emph{not} depend on the particular Riemann solver, while the force (with Eq. (\ref{eq:hllAvgDensity})) does.  
For a different Riemann solver, Eq. (\ref{eq:hllAvgDensity}) should be replaced with the face density consistent with that solver.  

\section{Simulations of SASI-induced magnetic field amplification}
\label{sec:results}

In this section we describe high-resolution MHD simulations with \genasis, exploring the capacity of nonlinear SASI-driven flows to amplify magnetic fields.  
With an idealized model we seek to investigate the character of SASI-induced magnetic field amplification.  
Interesting questions to explore include: does the presence of magnetic fields impact the evolution of the SASI?, what are the implications for observables associated with CCSNe, in particular PNS magnetization?  
However, with idealized simulations, we can only begin to scratch the surface of these complicated topics.  

\begin{wrapfigure}{r}{2.425in}
  \includegraphics[angle=00,height=2.297in,width=2.4in]{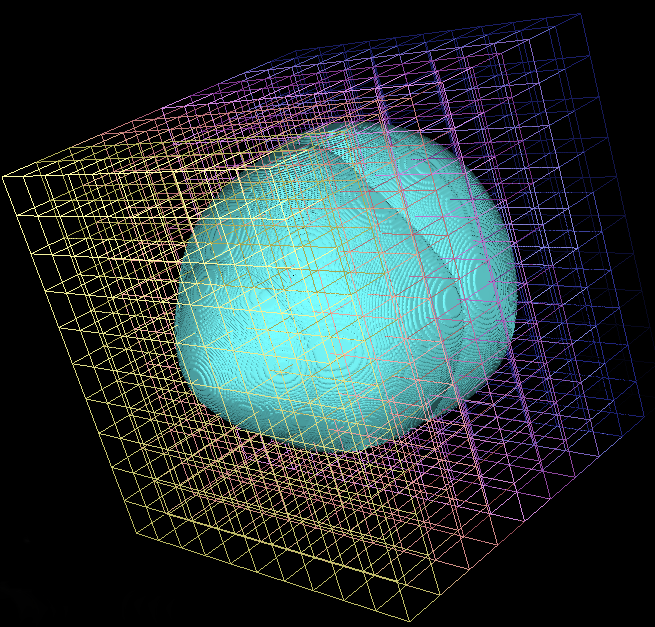}
  \caption{Domain composition from a 3D MHD-SASI simulation.}
  \label{fig:domains}
\end{wrapfigure}

\subsection{Model setup and parallel execution}
\label{sec:setup}

Our initial conditions follow closely the adiabatic setup described in \cite{blondin_etal_2003} and \cite{blondinMezzacappa_2007} (see also \cite{endeve_etal_2010}): a spherical stationary accretion shock is placed at a radius $\rShock=200$~km.  
A highly supersonic flow is nearly free-falling towards the shock for $r>\rShock$.  
Between the shock and the PNS the flow settles subsonically---obeying the Bernoulli equation $u_{r}^{2}/2+\gamma P/[\rho(\gamma-1)]-GM/r=0$---in nearly hydrostatic equilibrium.  
Matter is allowed to flow through an inner boundary placed at $\rPNS=40$~km.  
The ratio of specific heats is set to $\gamma=4/3$, the mass of the central object is held fixed at $M=1.2~M_{\odot}$, and the accretion rate ahead of the shock $\dot{M}=0.36~M_{\odot}\mbox{ s}^{-1}$ is also held fixed during the simulations.  
A radial magnetic field is superimposed on the initial condition: $B_{r}=\mbox{sign}(\cos\theta)\times B_{0}(\rPNS/r)^{2}$, where $B_{0}$ is the magnetic field strength at the surface of the PNS.  
We initiate the SASI with random pressure perturbations in the post-shock flow.  

\begin{wrapfigure}{l}{2.825in}
 \includegraphics[angle=00,height=2.58in,width=2.8in]{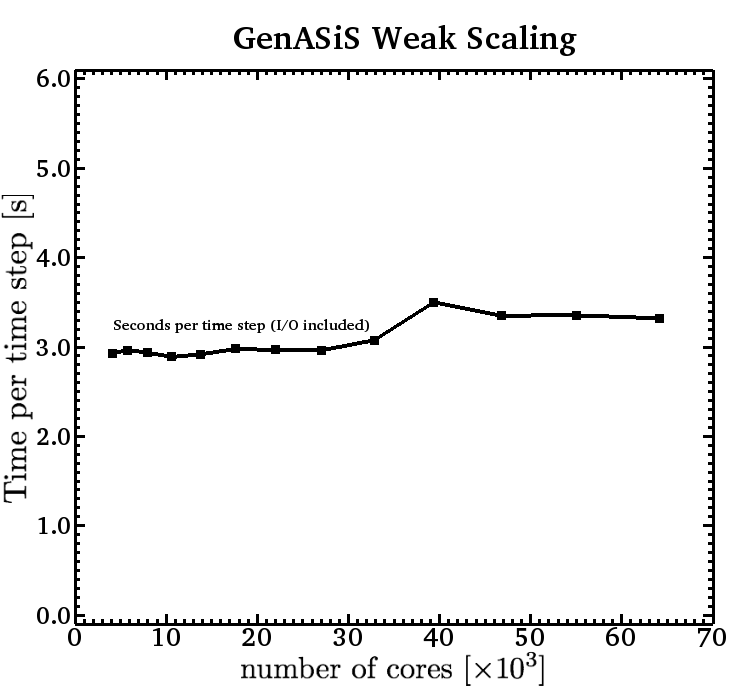}
    \caption{Weak scaling from a high-resolution 3D MHD-SASI simulation.}
    \label{fig:scaling}
\end{wrapfigure}

The MHD solver is parallelized using the Message Passing Interface (MPI), and the computational domain is subdivided into blocks containing an equal number of zones, which are distributed among MPI processes (see Figure \ref{fig:domains}).  
During the simulations we keep the number of zones per block (MPI process) fixed to $32^{3}$.  
To conserve computational resources the simulations are initiated in a relatively small computational domain with sides $L_{\mbox{\tiny min}}=600$~km covered by 512 zones ($\Delta x\approx$1.17~km).  
Once the SASI evolves into the nonlinear regime the volume encompassed by the shock grows, and the shock eventually interacts with the domain boundaries.  
When this happens, we expand the computational domain by adding a layer of $32^{3}$-zones blocks (i.e., we add 64 zones in each coordinate direction) and restart the simulation from the last checkpoint written before the shock interacted with the boundary.  
We repeat this process, and run our simulations until the shock interacts with the boundary of the largest computational box $L_{\mbox{\tiny max}}=1500$~km, or the simulation time reaches $t=1100$~ms, whichever occurs first.  
Since we keep $\Delta x$ fixed during the simulations, the largest computational domain is covered by 1280 zones in each spatial dimension.  
The weak scaling plot in Figure \ref{fig:scaling} shows the wall-time per time step, which remains reasonably constant as the computational domain expands to accommodate the growing shock volume.  
During a run, we write simulation output for analysis and visualization every 2~ms of physical time, resulting in tens of Terabytes of data from each model.  

\subsection{Simulation results}

\begin{figure}[h]
  \includegraphics[height=4.14in,width=4.0in]{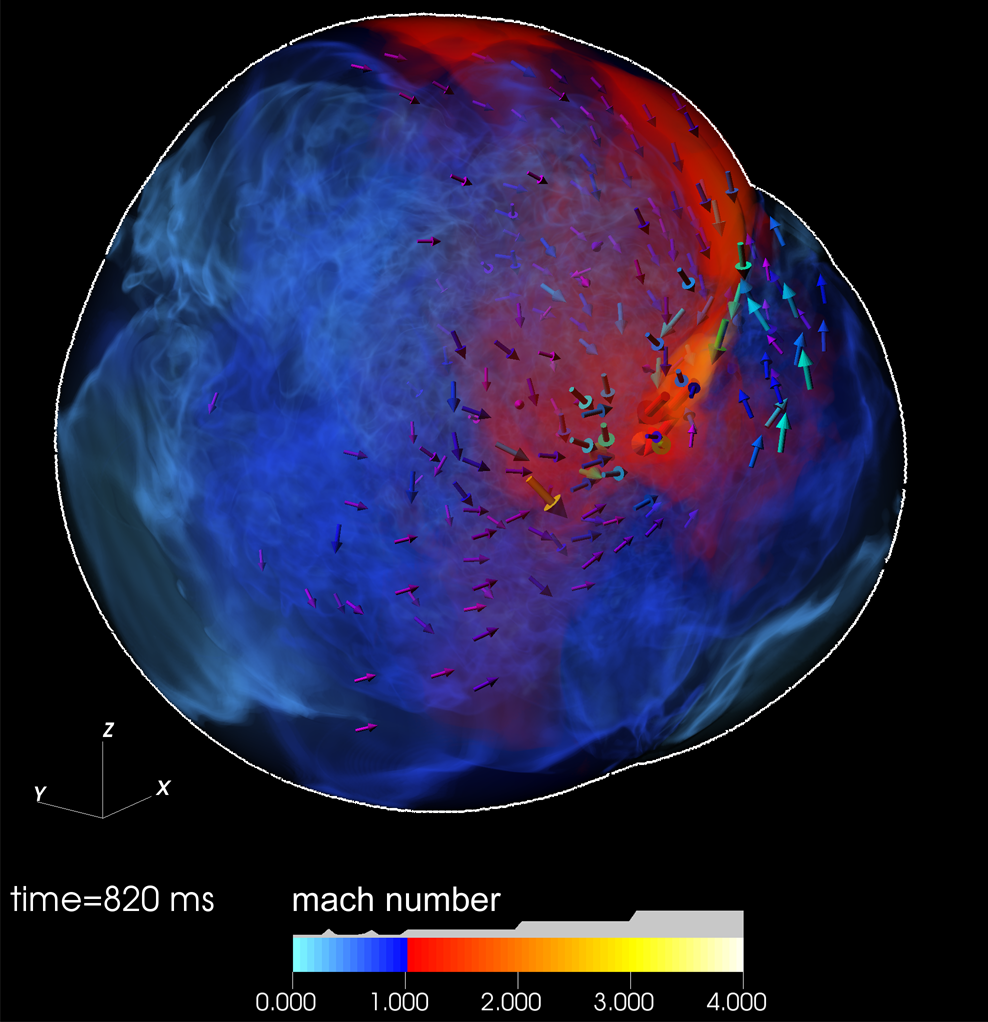}\hspace{0.1in}
  \begin{minipage}[b]{2.15in}
    \caption{Volume rendering of the Mach number $\mbox{Ma}=|\vect{u}|/c_{\mbox{\tiny s}}$ with velocity vectors ($|\vect{u}|\ge10^{4}$~km s$^{-1}$) superimposed.  
    Angular momentum redistribution has occurred as a result of the SASI, and the presence of counterrotating flows is apparent.  
    The shock triple-point \cite{blondinMezzacappa_2007} (a line segment extending across the shock surface), positioned to the upper right, is visible as the kink in the shock surface (white contour).  
    It connects the pre-shock accretion flow and the two counterrotating post-shock flows, and moves on the shock surface in the counterclockwise direction.  
    A layer of sheared flows extends from the triple-point and results in post-shock vorticity generation.  
    Image courtesy of Ross Toedte, ORNL.  \label{fig:machNumber}}
  \end{minipage}
\end{figure}

We have carried out simulations with varying initial magnetic field: $B_{0}=1\times10^{10}$~G (model $\model{10}$ or ``weak-field model''), $B_{0}=1\times10^{12}$~G (model $\model{12}$), and $B_{0}=1\times10^{13}$~G (model $\model{13}$ or ``strong-field model'').  
The initial field can be considered weak in all the models in the sense that the magnetic energy density is small compared to the kinetic and internal energy densities.  

Upon perturbation and nonlinear development, all models eventually exhibit typical spiral mode flows. 
This is consistent with \cite{blondinMezzacappa_2007}, who found the spiral mode to dominate the late-time evolution independent of the initial perturbation. 
Figure \ref{fig:machNumber} illustrates the post-shock flows associated with the nonlinear spiral SASI mode.  
(The rendering is created from a snapshot of the strong-field model at $t=820$~ms, but the hydrodynamic developments exhibited by this model are typical of all the computed models.)  

The pre-shock accretion flow impinges on the shock at an oblique angle due to the aspherical shock and its off-center position.  
The sizable tangential velocity component (relative to the shock surface), which is preserved across the shock, leads to supersonic post-shock flows ahead of (and directed towards) the triple-point.  
Supersonic shear flows are directed down towards the PNS and result in vorticity generation.  
Vorticity is distributed in a large fraction of the post-shock volume during the operation of the spiral SASI mode.  
Strongly forced accretion-driven turbulence develops as a result of the SASI, and the post-shock flow becomes roughly divided into a supersonic (driving) component and a subsonic (volume-filling) turbulent component.  
These hydrodynamic developments result in turbulent magnetic field amplification \cite{endeve_etal_2010,endeve_etal_2012}.  

\begin{figure}
  \includegraphics[angle=00,height=3.56in,width=6.3in]{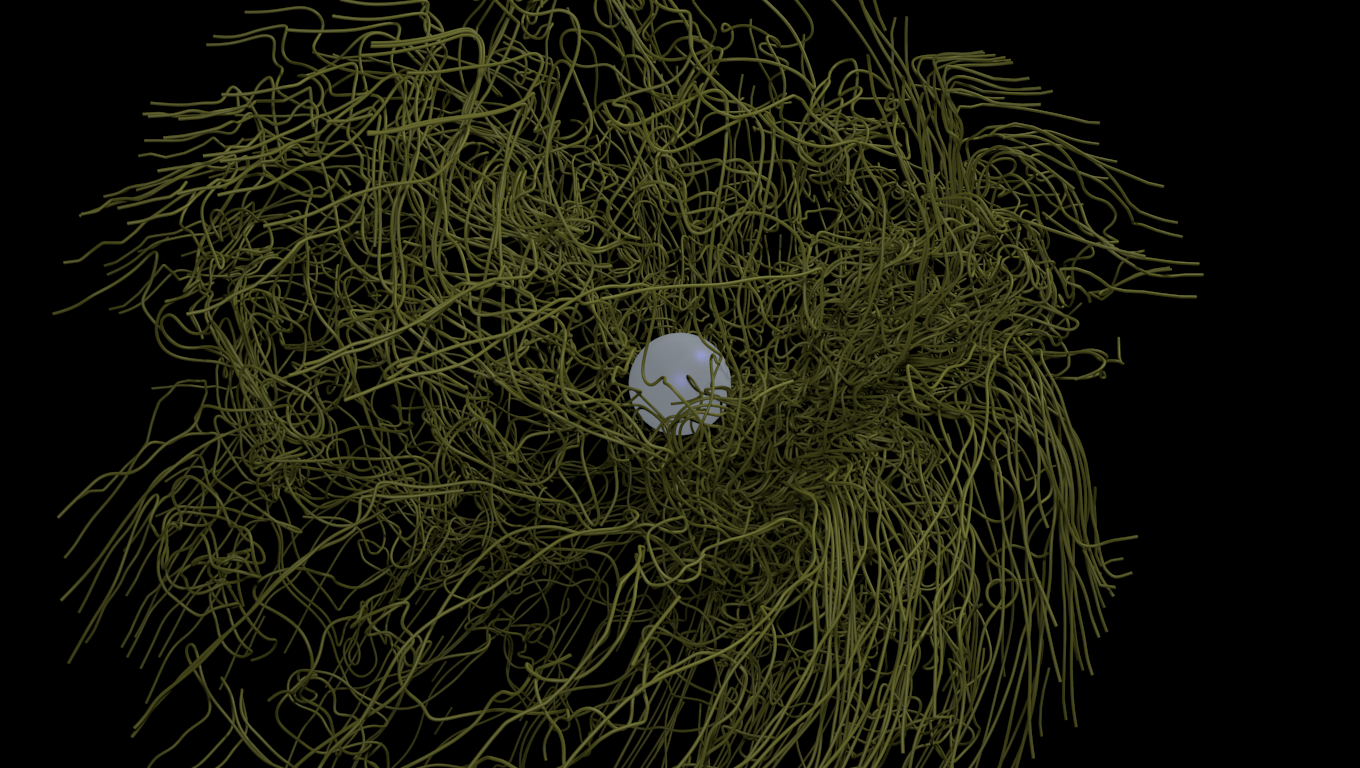}
    \caption{Streamlines tracing magnetic field lines in an MHD-SASI simulation during the nonlinear stage.  Streamlines are randomly seeded on the shock surface.  Image courtesy of Dave Pugmire, ORNL.  (An animation from which this image is taken can be viewed at {\small\tt http://events.cels.anl.gov/scidac11/visualization-night/visualization-night-winners/}.) \label{fig:streamlines}}
\end{figure}

The magnetic field amplifies in response to the hydrodynamics developments, and the magnetic energy becomes concentrated in thin intense magnetic flux tubes (or ropes).  
The volume rendering in Figure \ref{fig:streamlines} shows streamlines tracing out the magnetic field below the shock and reveals a complicated magnetic field geometry.  
The magnetic field is ``frozen-in" to the fluid in the ideal MHD limit.  
The large scale flows associated with the SASI result in relatively straight field lines, while the turbulent flows are responsible for a more ``chaotic" structure.  

\begin{figure}
  \includegraphics[angle=00,height=6.3in,width=6.3in]{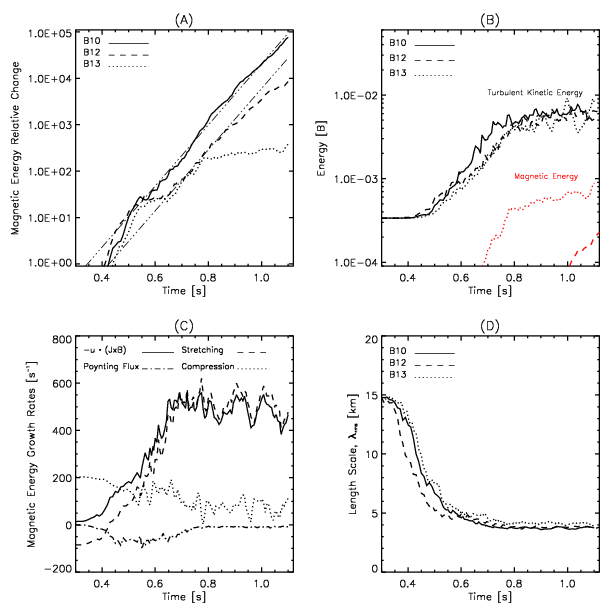}
    \caption{Summary of simulation results.  Panel (A): magnetic energy versus time; panel (B): turbulent kinetic energy versus time; panel (C): growth rates due to various terms in Eq. (\ref{eq:magneticEnergyEquation}) versus time for model $\model{10}$; and panel (D): magnetic rms scale (average flux rope thickness) versus time.  See text for further details and discussion.  ($1~\mbox{B}\equiv10^{51}$~erg.)  \label{fig:summaryOfResults}}
\end{figure}

Figure \ref{fig:summaryOfResults} contains further details from the numerical simulations.  
In Panel (A) we plot the relative change in total magnetic energy below the shock versus time for all models: $\model{10}$ (\full), $\model{12}$ (\dashed), and $\model{13}$ (\dotted).  
(The change is scaled to the initial magnetic energy for easy comparison across the models.)  
After an initial spurt, all the models experience an early period of exponential magnetic energy growth with essentially the same growth rate (cf. the temporal window from $650$~ms to $780$~ms).  
Exponential growth is typical during the kinematic regime of a turbulent dynamo, in which the magnetic field's back-reaction on the fluid is negligible (e.g., \cite{brandenburgSubramanian_2005}), and the growth time is roughly the turnover time of the turbulent eddies $\tau_{\mbox{\tiny eddy}}$.  
The magnetic energy in the weak-field model grows exponentially at a nearly constant rate, with growth time $\tau\approx66$~ms (dash-dotted reference lines), until the end of the simulation ($t=1100$~ms), and receives a total boost of about five orders of magnitude ($E_{\mbox{\tiny m}}\approx1.8\times10^{-7}$~B).  
In the model with $B_{0}=1\times10^{12}$~G ($\model{12}$), $E_{\mbox{\tiny m}}$ also grows steadily until the end of the run.  
The magnetic energy in this model receives a boost of four orders of magnitude ($E_{\mbox{\tiny m}}\approx2.3\times10^{-4}$~B).  
It grows initially at the same rate as the weak-field model, but the growth rate tapers off at later times ($t>900$~ms).  
The strong-field model ($\model{13}$, $B_{0}=1\times10^{13}$~G) exhibits exponential magnetic energy growth ($\tau\approx66$~ms) early on.  
Then, around $t\approx780$~ms, the growth rate drops almost discontinuously, and $E_{\mbox{\tiny mag}}$ grows by only about $50\%$ for the remainder of the simulation.  
Model $\model{13}$ receives a total boost in magnetic energy of about a factor of 300 ($E_{\mbox{\tiny m}}\approx8.9\times10^{-4}$~B).  
The abrupt drop in the magnetic energy growth rate observed in the strong-field model occurs when magnetic fields become dynamically important: concentrations of high $\beta_{\mbox{\tiny kin}}^{-1}(\equiv v_{\mbox{\tiny A}}^{2}/|\vect{u}|^{2})$, which exceeds unity in localized regions, are scattered throughout the shock volume.  
About $50\%$ of the total magnetic energy in model $\model{13}$ resides in regions where $\beta_{\mbox{\tiny kin}}^{-1}\ge10^{-1}$, and these magnetic fields occupy less than $10\%$ of the total shock volume.  
Regions with $v_{\mbox{\tiny A}}\gtrsim|\vect{u}|$ appear and disappear in a highly intermittent manner during this stage.  

The magnetic energy grows at the expense of turbulent kinetic energy below the shock.  
The turbulent kinetic energy $\EkinTurb=\int_{k_{\mbox{\tiny T}}}^{\infty}\hat{e}_{\mbox{\tiny k}}(k)\,dk$ is plotted in Panel (B).  
It is defined as flows varying on scales with wavenumber $k\ge\kTurb=2\pi/\lambdaTurb$, where $\lambdaTurb\approx30$~km.  
(The spectral kinetic energy density $\hat{e}_{\mbox{\tiny k}}(k)$ is obtained from Fourier transforms of components of the kinetic energy.)  
The particular choice for $\kTurb$ is motivated by several factors, including (1) magnetic field amplification occurs mostly on spatial scales with $k>\kTurb$ (from magnetic energy spectra we obtain a characteristic magnetic field scale $\lambdaMag\approx20$~km), and (2) the Taylor microscale $(\langle u^{2}\rangle/\langle |\curl{\vect{u}}|^{2}\rangle)^{1/2}$, which measures the average size of turbulent eddies, is comparable to $\lambdaTurb$.  

The turbulent kinetic energy grows rapidly during the ramp-up of the SASI and reaches a saturation level.  
After saturation, the time-averaged turbulent kinetic energy (averaged over the time interval from 800~ms to 1100~ms) in the respective models are found to be $\timeAverage{\EkinTurb}{0.8}{1.1}=5.90\times10^{-3}$~B ($\model{10}$), $5.48\times10^{-3}$~B ($\model{12}$), and $5.11\times10^{-3}$~B ($\model{13}$).  
(The total kinetic energy below the shock $E_{\mbox{\tiny k}}$ is about an order of magnitude larger during this stage.)  
We note that $\EkinTurb$ in model $\model{12}$ and model $\model{13}$ is reduced (relative to the weak-field model) by $\sim7\%$ ($4.2\times10^{-4}$~B) and $\sim13\%$ ($7.9\times10^{-4}$~B), respectively.  
The reduction in turbulent energy is comparable to the gain in magnetic energy in the respective models.  
Although there is a measurable reduction in the turbulent kinetic energy in the models attaining stronger $B$-fields, we find no impact of magnetic fields on large scale flows and SASI evolution.  
(The turbulent rms velocity $u_{\mbox{\tiny rms}}$ inferred from these models exceeds 4000~km~s$^{-1}$, and the associated eddy turnover time is $\tau_{\mbox{\tiny eddy}}=\bar{\lambda}_{\mbox{\tiny m}}/u_{\mbox{\tiny rms}}\approx5$~ms.)  

The saturation level for the turbulent kinetic energy 
may serve as an upper limit on the magnetic energy attainable in these simulations.  
We note that the magnetic energy in model $\model{13}$ (dotted red line in Panel (B)) seems to saturate at about $10\%$ of the turbulent kinetic energy.  
The magnetic fields in this model become dynamically significant, but are also influenced by numerical dissipation during the evolution (see below).  
The fluid in a developing CCSN is an excellent electrical conductor and magnetic energy dissipation is probably not important on the explosion time scale.  
Thus, the magnetic energy may possibly grow beyond the levels seen in our simulations, but probably not above the turbulent kinetic energy.  

Assuming a non-ideal electric field $-(\vect{u}\times\vect{B})+\eta\vect{J}$ with scalar resistivity $\eta$, the evolution equation for the magnetic energy density is easily obtained by dotting Eq. (\ref{eq:faradayLaw}) with $\vect{B}$:
\begin{eqnarray}
  \pderiv{e_{\mbox{\tiny m}}}{t}
  &=&
  \vect{B}\cdot
  \left[
    \left(\vect{B}\cdot\nabla\right)\vect{u}
    -\left(\vect{u}\cdot\nabla\right)\vect{B}
    -\vect{B}\left(\divergence{\vect{u}}\right)
    -\curl{\left(\eta\vect{J}\right)}
  \right] \nonumber \\
  &=&
  -\divergence{\vect{P}}
  -\vect{u}\cdot\left(\vect{J}\times\vect{B}\right)
  -\vect{B}\cdot\curl{\left(\eta\vect{J}\right)},
  \label{eq:magneticEnergyEquation}
\end{eqnarray}
where $\vect{P}=[\vect{u}(\vect{B}\cdot\vect{B})-\vect{B}(\vect{B}\cdot\vect{u})]$ and $\vect{J}=\curl{\vect{B}}$.  
(We use scalar resistivity here for sake of simplicity of the discussion, but it is also appropriate due to the similarity between the last two terms in Eq. (\ref{eq:HLLField}) and $\eta J_{z}$.)  
The first and third term on the right-hand-side of Eq. (\ref{eq:magneticEnergyEquation}) (first line) describe magnetic field amplification due to stretching and compression, respectively.  
Magnetic energy growth rates due to compression (\dotted) and stretching (\dashed) are plotted versus time for model $\model{10}$ in Panel (C) of Figure \ref{fig:summaryOfResults}.  
Stretching is the dominant magnetic field amplification mechanism in the nonlinear regime ($t\gtrsim650$).  
In a turbulent flow, the separation between two (initially adjacent) fluid elements grows exponentially with time.  
If the fluid elements are connected by a weak magnetic field the frozen-in condition of ideal MHD results in stretching, and thereby strengthening, of the magnetic field and an increase in the magnetic energy \cite{ott_1998}.  

The terms on the right-hand-side of Eq. (\ref{eq:magneticEnergyEquation}) (second line) represent Poynting fluxes, work done against the Lorentz force ($W_{\mbox{\tiny L}}$) and magnetic energy decay due to resistive dissipation ($-Q_{\mbox{\tiny J}}$), respectively.  
Kinetic energy is converted into magnetic energy if $W_{\mbox{\tiny L}}>0$.  
It is apparent from Eq. (\ref{eq:magneticEnergyEquation}) that the magnetic energy growth rate is due to the sum of individual rates from accretion of magnetic energy (Poynting flux) through the surface enclosing the PNS ($\partial V_{\mbox{\tiny PNS}}$), work done against the Lorentz force, and resistive dissipation.  
The Poynting flux through $\partial V_{\mbox{\tiny PNS}}$ and resistive dissipation generally result in decay of the magnetic energy in the computational domain.  
The decay must be overcome by the Lorentz work term in order for the magnetic energy to increase.  
(The magnetic Reynolds number is $R_{\mbox{\tiny m}}=|W_{\mbox{\tiny L}}|/|Q_{\mbox{\tiny J}}|$.)  
Growth rates due to work done against the Lorentz force (\full) and losses due to the Poynting flux through $\partial V_{\mbox{\tiny PNS}}$ (\chain) are plotted in Figure \ref{fig:summaryOfResults} (Panel (C)).  
Evidently, the growth due to $W_{\mbox{\tiny L}}$ greatly exceeds that due to Poynting losses through $\partial V_{\mbox{\tiny PNS}}$.  
Moreover, the growth rate is $\approx 480$~s$^{-1}$, which implies a magnetic energy growth time of about $2$~ms (comparable to $\tau_{\mbox{\tiny eddy}}$!), as opposed to the $66$~ms seen in Panel (A).  
This discrepancy in growth rates is due to numerical dissipation: the field evolves on small spatial scales in the turbulent flows and the dissipative terms in Eq. (\ref{eq:HLLField}) become nontrivial and affect the overall growth rate.  
This suggests that the magnetic energy may grow on a \emph{millisecond time scale} in SASI-driven turbulence with high $R_{\mbox{\tiny m}}$.  

\begin{wrapfigure}{r}{3.025in}
 \includegraphics[angle=00,height=3.0in,width=3.0in]{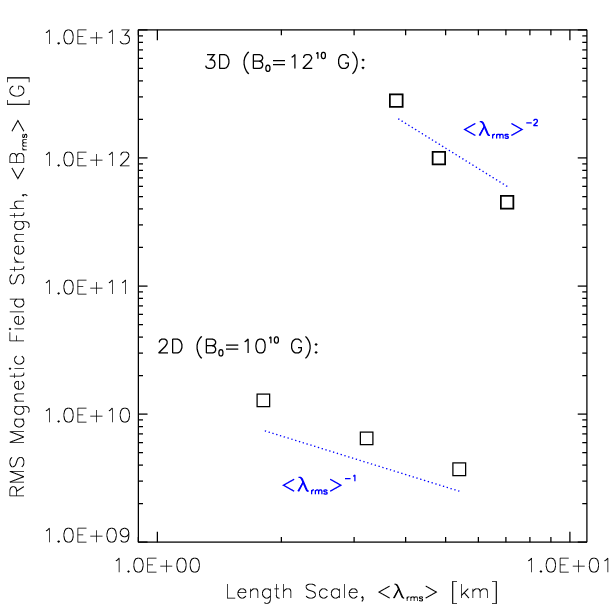}
    \caption{Plot of rms magnetic field $B_{\mbox{\tiny rms}}$ versus $\lambda_{\mbox{\tiny rms}}$.  Both quantities are averaged over an extended period in the nonlinear evolution.}
    \label{fig:fieldVSScale}
\end{wrapfigure}

The effect of numerical dissipation is illustrated in the evolution of magnetic flux tubes.  
In Panel (D) of Figure \ref{fig:summaryOfResults} we plot the average flux tube thickness $\lambda_{\mbox{\tiny rms}}=(\langle B^{2}\rangle/\langle|\gradient{B}|^{2}\rangle)^{1/2}$ versus time for all the models.  
During the ramp-up of the SASI, $\lambda_{\mbox{\tiny rms}}$ decreases rapidly until it reaches about $4$~km ($\sim3\,\Delta x$) and stays relatively constant thereafter.  
The decrease in flux rope thickness is halted by (numerical) resistive dissipation when $\lambda_{\mbox{\tiny rms}}$ reaches the resistive scale (a few$\times\Delta x$).  
In an environment with near perfect conductivity (i.e., the CCSN environment) we expect the field amplification to be halted by dynamical interactions with the fluid before the resistive scale is reached \cite{thompsonDuncan_1993}.  
In our large-scale simulations, finite spatial resolution and associated numerical dissipation plays a decisive role for the magnetic field evolution: the attainable field strength \emph{and} the magnetic energy growth rate are underestimated.  
By investigating individual terms in the induction equation during the late-time phase of the MHD-SASI simulations, we have verified that the dissipative terms in Eq. (\ref{eq:HLLField}) result in magnetic energy decay with a decay time comparable (but smaller in magnitude) to the growth time due to $W_{\mbox{\tiny L}}$.  

The geometric concept of magnetic flux ropes is further illustrated in Figure \ref{fig:fieldVSScale}, where we plot the time-averaged rms magnetic field $\langle B_{\mbox{\tiny rms}}\rangle$ versus time-averaged magnetic rms scale $\langle\lambda_{\mbox{\tiny rms}}\rangle$ from models where the spatial resolution has been varied ($\Delta x\approx2.34, 1.56, \mbox{ and } 1.17$~km).  
Higher spatial resolution accommodates the development of thinner flux ropes.  
The decreased cross-sectional area of a flux rope implies an increased field strength.  
In Figure \ref{fig:fieldVSScale}, the increase in field strength is in direct proportion to the decrease in the cross-sectional area afforded by higher spatial resolution.  
Also, the slope observed in the 3D simulations is steeper than that observed in 2D (axisymmetric) simulations \cite{endeve_etal_2010}.  
(We also observe that the exponential growth rate increases with increasing spatial resolution.)  

The increase in magnetic energy in the volume occupied by the PNS due to Poynting flux through $\partial V_{\mbox{\tiny PNS}}$ is $1.1\times10^{44}$~erg ($\model{10}$), $3.2\times10^{47}$~erg ($\model{12}$), and $4.5\times10^{48}$~erg ($\model{13}$).  
The results form models $\model{12}$ and $\model{13}$ imply PNS rms magnetic fields exceeding $10^{14}$~G.  
We expect the degree of PNS magnetization to be insensitive to the initial field in realistic settings with high $R_{\mbox{\tiny m}}$ where dissipative processes can be ignored.  

\section{Summary and discussion}

We have presented the initial implementation of a second-order finite volume MHD scheme in \genasis.  
Using the MHD capabilities, we have performed 3D simulations of the evolution and amplification of magnetic fields in SASI-driven flows.  
The simulations are initiated from a configuration which resembles the early stalled shock phase in a CCSN, albeit with simplified physics that excludes critical components of a supernova model (e.g., neutrino transport, self-gravity, and the PNS itself).
On the other hand, our simulations are computed with a spatial resolution that is currently inaccessible to state-of-the-art supernova models in three spatial dimensions, and they may therefore provide valuable insight into MHD developments in CCSNe.  

Flows associated with the spiral SASI mode result in vigorous turbulence below the shock ($u_{\mbox{\tiny rms}}\approx4000$~km~s$^{-1}$).  
The turbulence amplifies magnetic fields by stretching, and the magnetic energy grows exponentially with time as long as the kinematic regime obtains.  
The resulting magnetic fields display a highly intermittent flux rope structure.  
In the strong-field model, the magnetic energy saturates when the associated energy density becomes comparable to the kinetic energy density (i.e., $v_{\mbox{\tiny A}}\sim|\vect{u}|$) in localized regions of the flow.  
The subsequent magnetic field evolution remains highly dynamic: strong fields emerge, are advected with the flow, are temporarily weakened, and then reemerge in a seemingly stochastic manner.  

The presence of amplified magnetic fields does not result in noticeable effects on the global shock dynamics in our simulations, and this may be understood from considerations of the energetics: the turbulent kinetic energy---which powers SASI-driven field amplification---only amounts to about $10\%$ of the total kinetic energy below the shock ($E_{\mbox{\tiny k}}\sim5\times10^{-2}$~B), and is not significant to the explosion ($\sim1$~B).  
These observations alone suggest a rather passive role of magnetic fields in the overall dynamics of non-rotating CCSNe.  

Our simulations further suggest that the SASI may play a role in determining the strength of the magnetic field in proto-neutron stars and young pulsars.  
We estimate that the magnetic energy accumulated on the PNS may account for magnetic field strengths exceeding $10^{14}$~G.  

The evolution of magnetic fields in our simulations remains sensitive to the spatial resolution.  
The magnetic energy attained and the rate at which the magnetic energy grows increase with increasing grid resolution.  
Data extracted from our simulations suggest that the magnetic energy may grow exponentially on a millisecond timescale ($\sim\tau_{\mbox{\tiny eddy}}$) under more realistic physical conditions (i.e., huge magnetic Reynolds numbers), as opposed to the $\sim50$-$60$~ms timescale measured directly in our runs.  
Model $\model{13}$ attains dynamically relevant magnetic fields, which are also subject to significant numerical dissipation.  
The MHD evolution displayed by this model during the saturated state is probably spurious and the dynamical impact of the $B$-fields may be underestimated.  
We therefore caution that the uncertainties associated with the sensitivity to numerical resolution prevents us from completely dismissing magnetic fields as unimportant to the explosion dynamics of weakly rotating progenitors.  


In summary, we conclude from our simulations that magnetic fields in CCSNe may be amplified exponentially by turbulence driven by the spiral SASI mode.  
Details on the impact of SASI-amplified magnetic fields on explosion dynamics remain unclear, but on energetic grounds alone the role of these magnetic fields is likely sub-dominant.  
The simulations further suggest that small-scale PNS magnetic fields in the $10^{14}-10^{15}$~G range may be formed, which may be sufficient to power some of the energetic activity that define AXPs and SGRs.  

\ack

This research was supported by the Office of Advanced Scientific Computing Research and the Office of Nuclear Physics, U.S. Department of Energy.  
This research used resources of the Oak Ridge Leadership Computing Facility at the Oak Ridge National Laboratory provided through the INCITE program.

\end{document}